\documentclass[12pt]{article}
\usepackage{graphicx,psfrag,cite}

\newcommand{\bea}{\begin{eqnarray}}
\newcommand{\eea}{\end{eqnarray}}
\newcommand{\simgt}{\hbox{ \raise3pt\hbox to 0pt{$>$}\raise-3pt\hbox{$\sim$} }}
\newcommand{\simlt}{\hbox{ \raise3pt\hbox to 0pt{$<$}\raise-3pt\hbox{$\sim$} }}

\begin{document}
\begin{titlepage}
\title{Perturbative QCD Potential, Renormalon Cancellation
and Phenomenological Potentials \vspace{2cm}}
\author{S.~Recksiegel$^1$ and Y.~Sumino$^2$
\\
\\ $^1$ Theory Group, KEK\\
Tsukuba, Ibaraki, 305-0801 Japan\\
\\ $^2$ Department of Physics, Tohoku University\\
Sendai, 980-8578 Japan
}
\date{}
\maketitle
\thispagestyle{empty}
\vspace{-5truein}
\begin{flushright}
{\bf hep-ph/0109122}\\
{\bf TU--629}\\
{\bf KEK--TH--783}\\
{\bf September 2001}
\end{flushright}
\vspace{4.5truein}
\begin{abstract}
\noindent
{\small
We examine the total energy 
$E_{\rm tot}(r) = 2 m_{{\rm pole}} + V_{\rm QCD}(r)$
of the $b\bar{b}$ system within perturbative QCD
up to ${\cal O}(\alpha_S^3)$.
We extend a previous analysis by incorporating effects of
the non-zero charm-quark mass in loops.
We find that, once the renormalon cancellation is performed, 
$E_{\rm tot}(r)$ agrees well with 
typical phenomenological potentials for heavy
quarkonia at distances
$0.5~{\rm GeV}^{-1} \simlt r \simlt 3~{\rm GeV}^{-1}$.
We also examine the perturbative predictions for
$E_{\rm tot}(r)$ of other heavy quarkonium systems.
Whenever stable predictions are obtained, they agree with
each other up to an $r$-independent constant.
}
\end{abstract}
\vfil

\end{titlepage}
  
\section{Introduction}

Heavy quarkonium systems, such as bottomonium and charmonium, provide
an important 
testing ground for theoretical studies on the dynamics of QCD boundstates.
For decades, theoretical methods for investigating these systems
have followed paths different from those used for studying the
QED boundstates such as positronium or the hydrogen atom.
The reason is that the non-relativistic boundstate theory
based on perturbative QCD, by itself, was not successful for
describing the heavy quarkonium states.
The most serious problem has been the fact that perturbative QCD
does not reproduce the shape of the static QCD potential $V_{\rm QCD}(r)$,
which is considered to dictate a dominant part of the dynamics of
these system, in the relevant region
$0.5~{\rm GeV}^{-1} \simlt r \simlt 5~{\rm GeV}^{-1}$.

The theoretical method which has been widely used for analyses of
the heavy quarkonium states is the phenomenological potential-model
approach.
In this method, one assumes some simple form for an effective
(non-relativistic) Hamiltonian of the quarkonium system.
A phenomenological potential, which is close to the
above QCD potential conceptually, is introduced in the Hamiltonian.
One parameterises the potential and determines the parameters
such that various physical observables
of the quarkonia are reproduced in this model.
It turned out that phenomenological potentials determined in this way
have more or less similar shapes in the region
$0.5~{\rm GeV}^{-1} \simlt r \simlt 5~{\rm GeV}^{-1}$,
which may be represented typically by a Coulomb-plus-linear potential.
See e.g.\ Ref.~\cite{eq} for a recent analysis based on 
potential models.

On the other hand, theoretical
approaches, which have foundations on first principles,
have been developed.
Within the framework of Non-Relativistic QCD (NRQCD) or potential-NRQCD,
the Lagrangian of a heavy quarkonium 
system is given in a series expansion of some
small parameter (typically the quark velocity $v \sim \alpha_s$).
The Wilson coefficients of the effective theory are determined by
matching the theory to the full QCD theory either perturbatively or
non-perturbatively.
The QCD potential, as well as other (sub-leading)
potentials which enter the
effective theory, are determined from the results of lattice calculations
or from model calculations;
see \cite{bali} and references therein.
The lattice results, being first principle calculations, turn out to be 
consistent with phenomenologically determined potentials, but at the 
present stage they are still not precise enough.

Computations of the QCD potential within perturbative QCD also
made progress over time.
The full two-loop corrections with massless quark loops \cite{ps},
as well as non-zero mass effects in quark loops up to the same order
\cite{melles1,hoangmceff,melles2},
have been computed.
The perturbative expansion at $r \simgt 0.2~{\rm GeV}^{-1}$
revealed to be very poorly convergent, and also its 
shape deviates qualitatively
from an expected Coulomb-plus-linear form in the relevant range.
The poor convergence is considered to reflect
a non-trivial structure of the QCD vacuum.
Also, within the context of perturbative QCD, this behaviour
has been understood using the renormalon language \cite{al}.

Recently, there have been significant developments in
the non-relativistic boundstate theory based on perturbative
QCD.
Thanks to the calculations of higher-order corrections 
\cite{yn3,py,my,bcbv} and the understanding
of the leading renormalon cancellation \cite{renormalon1,renormalon2},
it became possible to predict accurately the physical observables of
the heavy quarkonium states (in particular, the bottomonium states)
within perturbative QCD 
\cite{pp,my,upsilonmass,hoangmceff,bsv1,paper1,rs,bsv2}.
An essential feature is that the prediction for the QCD potential,
after incorporating these developments, has become accurate, and that
it has reproduced a realistic shape of the potential in the relevant range.
Ref.~\cite{paper1} investigated this feature in particular.
There, the following aspects have been shown:
\begin{itemize}
\item[(1)] When the leading renormalon cancellation is incorporated, convergence
property of the total energy of the bottomonium system
$E_{\rm tot}(r)=2m_{b,{\rm pole}}+V_{\rm QCD}(r)$
improves drastically.
\item[(2)] For simplicity, two hypothetical cases, $m_c \to 0$ and
$m_c \to m_b$, have been examined;
a reliable theoretical prediction for $E_{\rm tot}(r)$ is
obtained at $r \simlt 3~{\rm GeV}^{-1}$, and the prediction 
agrees with typical phenomenological 
potentials in the range $0.5~{\rm GeV}^{-1} \simlt r \simlt 3~{\rm GeV}^{-1}$ 
within theoretical
uncertainties.
\item[(3)]The qualitative behaviour of $E_{\rm tot}(r)$ in the above range
can be understood as originating from an increase of the interquark force due
to the running of the coupling constant.
\end{itemize}

In this paper we extend the analysis of \cite{paper1}.
First, we incorporate the realistic value of the charm mass
in the calculation of $E_{\rm tot}(r)$.
Since we are interested in the range of $r$ not very different from the
charm-mass scale $1/m_c$, we should properly take into account the 
dependence on $m_c$ which enters through loop corrections.
In this way
we can compare predictions of perturbative QCD in the realistic case
with phenomenological potentials.
Secondly, we compute the total energies for quark-antiquark systems 
other than the $b\bar{b}$ system and compare them.
In \cite{paper1} the part of $E_{\rm tot}(r)$ independent of the external
quark masses has been examined using the interquark force.
Here we examine it in a different way.
Our purpose is to compare the QCD potential (or the corresponding
potential in the effective Hamiltonian of the quarkonium system),
which has been studied in various approaches so far, with
the prediction of perturbative QCD.
We anticipate that, by combining our results with the conventional studies,
we would be able to obtain a better understanding on
the dynamics of the heavy quarkonia from first principles.

In Sec.~2 we compute the total energy of the $b\bar{b}$ system,
incorporating non-zero charm mass effects:
in Sec.~2.1 we set up our formulas; in Sec.~2.2 numerical analyses
are given; in Sec.~2.3 we discuss uncertainties of our predictions.
In Sec.~3 the total energies of other systems are examined.
Conclusions are given in Sec.~4.
Appendices collect some formulas.

\section{Total Energy of the \boldmath{$b\bar{b}$} System}

\subsection{Definitions}
\label{definitions}

In the $\overline{\rm MS}$ scheme, it is appropriate to compute the total 
energy of the $b\bar{b}$ system 
in the theory which contains 5 flavours.
We are, however, interested in the total energy when the distance between 
$b$ and $\bar{b}$ is much larger than their Compton wavelength,
$r \gg 1/m_b$.
Therefore, we will rewrite the total energy in terms of the 4-flavour coupling
$\alpha_S^{(4)}(\mu)$ in order to realize decoupling of the $b$-quark to all
orders.

When we neglect the charm quark mass, the total energy is given by
\bea
E_{{\rm tot},{m}_c=0}^{b\bar{b}}(r) = 
2 m_{b,{\rm pole}} + V_{{\rm QCD},4}(r) .
\eea
The relation between the pole mass and the $\overline{\rm MS}$ mass
has been computed up to 3 loops in a full theory,
which contains $n_h$ heavy flavours and $n_l$ massless flavours \cite{mr}.
(The same relation was obtained numerically in \cite{chst} in a certain 
approximation.)
Setting $n_h=1$ and rewriting the relation in terms of the coupling of the
theory with $n_l=4$ massless flavours only, we find\footnote{
This relation coincides with Eq.(14) of \cite{mr}, which
is given numerically (indirectly through $\beta_0$ 
and $\beta_1$).
Note that,  in the other
formulas of \cite{mr}, the coupling of the full theory is used.
}
\bea
m_{b,{\rm pole}} = \overline{m}_b
\left\{ 1 + {4\over 3}\,  
{\alpha_S^{(4)}(\overline{m}_b)\over \pi} 
+   \left({\alpha_S^{(4)}(\overline{m}_b)\over \pi}\right)^2 d_1^{(4)} 
+  \left({\alpha_S^{(4)}(\overline{m}_b)\over \pi}\right)^3 d_2^{(4)} \right\}
\label{massrel},
\eea
where
$\overline{m}_i \equiv m_i^{\overline{\rm MS}}(m_i^{\overline{\rm MS}})$
denotes the $\overline{\rm MS}$ mass renormalised at the 
$\overline{\rm MS}$-mass scale.
The QCD potential of the theory with $n_l$ massless flavours only\footnote{
Due to the decoupling theorem, the perturbative 
QCD potential of the theory which contains
one heavy flavour (with mass $m$) and $n_l$ massless flavours
coincides with the potential in Eq.~(\ref{QCDpot}) up to
${\cal O}(\alpha_S^3)$ if we count $1/r = {\cal O} (\alpha_S m)$
and if we rewrite the coupling by the coupling of the theory with
$n_l$ massless flavours only.
}
is given, up to ${\cal O}(\alpha_S^3)$, by 
\bea
V_{{\rm QCD},n_l}(r) & = &
- \, \frac{4}{3} \frac{\alpha_S^{(n_l)}(\mu)}{r}
\Biggl[ \, 1 + 
\biggl( \frac{\alpha_S^{(n_l)}(\mu)}{4\pi} \biggr) 
  ( 2 \beta_0^{(n_l)} \ell + a_1^{(n_l)} )
\nonumber \\ && ~~~~~~~~~~~~~~
+
\biggl( \frac{\alpha_S^{(n_l)}(\mu)}{4\pi} \biggr)^2
  \left\{ \left( \beta^{(n_l)}_0 \right)^2 
\Bigl( 4 \ell^2  + \frac{\pi^2}{3} \Bigr)
  + 2 ( \beta^{(n_l)}_1 + 2 \beta^{(n_l)}_0 a_1^{(n_l)} ) \ell + a_2^{(n_l)} \right\}
\Biggr] ,
\nonumber\\
\label{QCDpot}
\eea
where 
$\ell = \log (\mu r) + \gamma_E $,
and $\beta_i$ denote the coefficients of the beta function
\bea
\beta_0^{(n_l)} = 11 - \frac{2}{3} n_l, 
~~~~~~
\beta_1^{(n_l)} = 102 - \frac{38}{3} n_l .
\eea
The constants $a_i$ and $d_i$ are given in Appendix A.

When we include the effects of the non-zero charm-quark mass,
the above formula is modified as follows:
\bea
E_{{\rm tot}}^{b\bar{b}}(r) = 
E_{{\rm tot},{m}_c=0}^{b\bar{b}}(r) +
2 \, \delta m_{b,{\rm pole}} + \delta V_{\rm QCD}(r) .
\eea
At ${\cal O}(\alpha_S^2)$ the non-zero charm mass correction to the pole mass,
$\delta m_{b,{\rm pole}}$, reads \cite{gray}
\bea
\delta m_{b,{\rm pole}}^{[2]} 
&=& {\overline{m}_b\over 3}
\left({\alpha_S^{(4)}({\overline{m}_b})\over \pi}\right)^2 \left[
\log^2(\xi) + {\pi^2\over 6} - \left(\log(\xi) + {3\over2}\right)\xi^2 \right. \nonumber\\
& &\quad +(1+\xi)(1+\xi^3)\left({\rm Li}_2(-\xi)-{1\over 2}\log^2(\xi)+\log(\xi)\log(1+\xi)
+{\pi^2\over 6}\right) \nonumber\\
& &\quad \left. 
+(1-\xi)(1-\xi^3)\left({\rm Li}_2(\xi)-{1\over 2}\log^2(\xi)+\log(\xi)\log(1-\xi)
-{\pi^2\over 3}\right)\right] ,
\label{dmpole1}
\eea
where $\xi = \overline{m}_c / \overline{m}_b$.
Its leading term in 
the limit $m_c \to 0$ (linear approximation) is given by
\bea
\delta m_{b,{\rm pole}}^{[2]}\Bigr|_{m_c \to 0} =
{(\alpha_S^{(4)}({\overline{m}_b}))^2\over 6} {\overline{m}_c} .
\eea
At ${\cal O}(\alpha_S^3)$, the complete expression of
$\delta m_{b,{\rm pole}}$ is not known; it has been computed only in the
linear approximation \cite{hoangmceff}:
\bea
\delta m_{b,{\rm pole}}^{[3]}\Bigr|_{m_c \to 0} 
&=& 
{(\alpha_S^{(4)}({\overline{m}_b}))^3\over \pi} {\overline{m}_c}
\left\{ {2\over 9} + {\beta_0^{(4)}\over 12}\left(
- 2\log ( \xi ) -4\log 2 +{14\over 3}\right) \right.\nonumber\\
& & \qquad \left. -{1\over 9}\left({59\over 15}+2 \log 2 \right) 
+ {19\over 9\pi}(f_1f_2 + b_1b_2)\right\} ,
\eea
where $f_2 = 0.470\pm0.005$, $b_2 = 1.120\pm0.010$,
$f_1=(\log A - \log b_2)/(\log f_2 -\log b_2)$, 
$b_1=(\log A - \log f_2)/(\log b_2 -\log f_2)$ and $\log A = 161/228 + 13\zeta_3/19 -\log 2$.
In \cite{hoangmceff} it has been argued that the use of the linear approximation
in both $\delta m_{b,{\rm pole}}^{[2]}$ and $\delta m_{b,{\rm pole}}^{[3]}$
is a slightly better approximation of the full result than to use
the exact $\delta m_{b,{\rm pole}}^{[2]}$ and the linear approximation
of $\delta m_{b,{\rm pole}}^{[3]}$.
It has been conjectured that the former approximation accounts for
the full correction 
$\delta m_{b,{\rm pole}}^{[2]} + \delta m_{b,{\rm pole}}^{[3]}$
with about 10\% accuracy, while
the latter approximation accounts for the full result with
about 20\% accuracy.
In \cite{bsv2}, the theoretical predictions for the bottomonium energy levels
turned out to be more stable when we used 
$\bigl(\delta m_{b,{\rm pole}}^{[2]}\bigr|_{m_c \to 0}\bigr) + 
\bigl(\delta m_{b,{\rm pole}}^{[3]}\bigr|_{m_c \to 0}\bigr)$ 
as compared to
$\delta m_{b,{\rm pole}}^{[2]} + 
\bigl(\delta m_{b,{\rm pole}}^{[3]}\bigr|_{m_c \to 0}\bigr)$. 

At ${\cal O}(\alpha_S^2)$ the non-zero charm mass correction to the QCD
potential,
$\delta V_{\rm QCD}$, is given in one-parameter integral form as 
\bea
\delta V_{\rm QCD}^{[2]}(r) = 
- \, \frac{4}{3} \,  \frac{\alpha_S^{(4)}(\mu)}{r}
\biggl( \frac{\alpha_S^{(4)}(\mu)}{3\pi} \biggr) 
\left[
\int_1^\infty dx \, f(x) \, e^{-2 \overline{m}_c r x} 
+ \Bigl( \log (\overline{m}_c r) + \gamma_E + \frac{5}{6} \Bigr)
\right] 
\eea
with
\bea
f(x) = \frac{\sqrt{x^2-1}}{x^2} \left( 1 + \frac{1}{2x^2} \right) .
\eea
At ${\cal O}(\alpha_S^3)$, the correction was computed 
first in momentum-space in \cite{melles1}.
The Fourier transform was performed\footnote{
The coordinate-space potential was studied using different 
integral representations in \cite{melles1}.
}
and $\delta V_{\rm QCD}^{[3]}(r)$ 
was obtained also in one-parameter integral form in 
\cite{hoangmceff,melles2}.
Both of the latter references contain misprints, however; for
completeness, we give a corrected formula
for $\delta V_{\rm QCD}^{[3]}(r)$ in Appendix B.\footnote{
The corrected formula has been acknowledged
by the author of \cite{hoangmceff}.
} 

Since the renormalon cancellation at each order of the
perturbative expansion
is realized only when we use the same coupling constant in expanding
$m_{b,{\rm pole}}$ and $V_{{\rm QCD},4}(r)$,
we rewrite $\alpha_S(\overline{m}_b)$ in terms of
$\alpha_S(\mu)$ using the renormalization-group evolution of the
coupling constant:
\bea
\alpha_S^{(n_l)}(\overline{m}_i)&=& \alpha_S^{(n_l)}(\mu) \left\{ 
1 +   \frac{\alpha_S^{(n_l)}(\mu)}{\pi}  \,\frac{\beta^{(n_l)}_0}{2} 
\log\left(\frac{\mu}{\overline{m}_i}\right) \right. \nonumber \\
& &\qquad\qquad\quad   \left.
+  \left( \frac{\alpha_S^{(n_l)}(\mu)}{\pi} \right)^2 \,
\Biggl[ \frac{\beta^{(n_l)\,2}_0}{4} \, \log^2 \left(\frac{\mu}{\overline{m}_i}\right)
+ \frac{\beta^{(n_l)}_1}{8} \log\left(\frac{\mu}{\overline{m}_i}\right) \Biggr] \right\}.
\label{alphams}
\eea
Furthermore, we also examine the total energy after re-expressing it
in terms of the 3-flavour coupling and compare it with the 4-flavour
coupling case.
This is because we are interested in the total energy in the range
$0.5~{\rm GeV}^{-1} \simlt r \simlt 5~{\rm GeV}^{-1}$,
where the charm quark may decouple as well.
We insert the relation \cite{lrv}
\bea 
\alpha_S^{(4)}(\mu) &=& \alpha_S^{(3)}(\mu)\left\{ 
1 + \frac{\alpha_S^{(3)}(\mu)}{3 \pi}  \, 
\log\left(\frac{\mu}{\overline{m}_c}\right) \right. \nonumber\\
& &\qquad\qquad\quad   \left.
+ \left( \frac{\alpha_S^{(3)}(\mu)}{\pi} \right)^2 \,
\Biggl[ {1\over 9} \, \log^2 \left(\frac{\mu}{\overline{m}_c}\right)
+ {19\over 12}\log\left(\frac{\mu}{\overline{m}_c}\right) - {11\over 72}\Biggr] \right\}
\label{alphams2}
\eea
into the above $E_{\rm tot}^{b\bar{b}}(r)$ 
and re-expand in $\alpha_S^{(3)}(\mu)$.
Thus, we will examine the series expansion of
$E^{b\bar{b}}_{\rm tot}
(r;\overline{m}_b,\overline{m}_c,\alpha_S^{(n_l)}(\mu))$ 
in $\alpha_S^{(n_l)}(\mu)$ up to 
${\cal O}((\alpha^{(n_l)}_S)^3)$ for $n_l =3$ and 4.

The obtained total energy depends on the scale $\mu$ due to 
truncation of the series at a finite order.
Following the prescriptions of \cite{paper1},
we will fix the scale $\mu$ in the two different ways 
described below:
\begin{enumerate}
\item
We fix the scale $\mu = \mu_1(r)$ by demanding stability 
of $E_{\rm tot}(r)$ against variation of 
the scale:
\bea
\left. \mu \frac{d}{d\mu} E_{\rm tot}(r;\overline{m}_i,\alpha_S(\mu))
\right|_{\mu = \mu_1(r)} = 0 .
\label{scalefix1}
\eea
\item
We fix the scale $\mu = \mu_2(r)$
on the minimum of the absolute value of the last known term 
[${\cal O}(\alpha_S^3)$ term] of $E_{\rm tot}(r)$:
\bea
\left. \mu \frac{d}{d\mu}
\Bigl[ E^{[3]}_{\rm tot}(r;\overline{m}_i,\alpha_S(\mu)) \Bigr]^2 \,
\right|_{\mu = \mu_2(r)} = 0 .
\label{scalefix2}
\eea
\end{enumerate}

\subsection{Numerical Analyses}
\label{numanal}

In this subsection we take the input value for the coupling constant as
$\alpha_S^{(5)}(M_Z)=0.1181 \pm 0.0020$ \cite{pdg}.
We evolve the coupling by solving
the 3-loop renormalization-group equation numerically and match it
to the 4- and 3-flavour couplings successively through the matching
condition \cite{lrv}.\footnote{
The matching scales are taken as $\overline{m}_b$ and $\overline{m}_c$,
respectively.
}
For the bottom- and charm-quark masses, we use the values
$\overline{m}_b=4.190^{-20}_{+19}$~GeV \cite{bsv2} and 
$\overline{m}_c = 1.243$~GeV \cite{bsv1}, respectively.
(For simplicity we do not change the value of $\overline{m}_c$
as a function of $\alpha_S^{(5)}(M_Z)$. 
This is justified, since 
the dependence of $E_{\rm tot}(r)$ on it is much smaller than other
theoretical uncertainties; see Sec.~\ref{errorest}.)
We compute both $\delta m_{b,{\rm pole}}^{[2]}$ and 
$\delta m_{b,{\rm pole}}^{[3]}$ in the linear approximation in
this subsection.

\begin{figure}
\begin{center}
\psfrag{XXX}{$\mu\,[{\rm GeV}]$}  
\psfrag{YYY}{$E_{\rm tot}\,[{\rm GeV}]$} 
\psfrag{ZZZ}{$|E_{\rm tot}^{[3]}|^2\,\,[{\rm GeV}^2]$}
\psfrag{Etot}{$E_{\rm tot}$}
\psfrag{Etot3}{$|E_{\rm tot}^{[3]}|^2$}
\includegraphics[width=13cm]{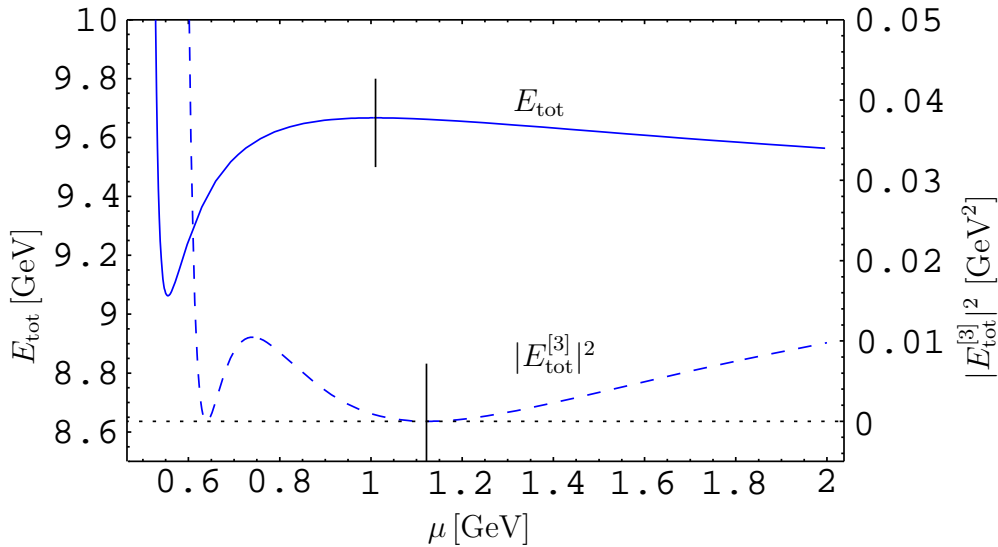}
\end{center}
\vspace*{-.5cm}
\caption{Determination of the scale: Total energy and 
$\alpha_s^3$-term of the total energy, the scales determined
with the respective prescription are marked with a short
vertical line. 
(We do not take the scales which are located close to the
infrared singularity of $\alpha_S^{(3)}(\mu)$.)
The distance $r$ is fixed to $2~{\rm GeV}^{-1}$; $\alpha_S^{(5)}(M_Z)=0.1181$.}
\label{findmu}
\end{figure}

\begin{figure}
\begin{center}
\psfrag{XXX}{$r\,[{\rm GeV}^{-1}]$}
\psfrag{YYY}{$E_{\rm tot}\,[{\rm GeV}]$}
\psfrag{3flavour}{3-flavour prescription}
\psfrag{4flavour}{4-flavour prescription}
\includegraphics[width=13cm]{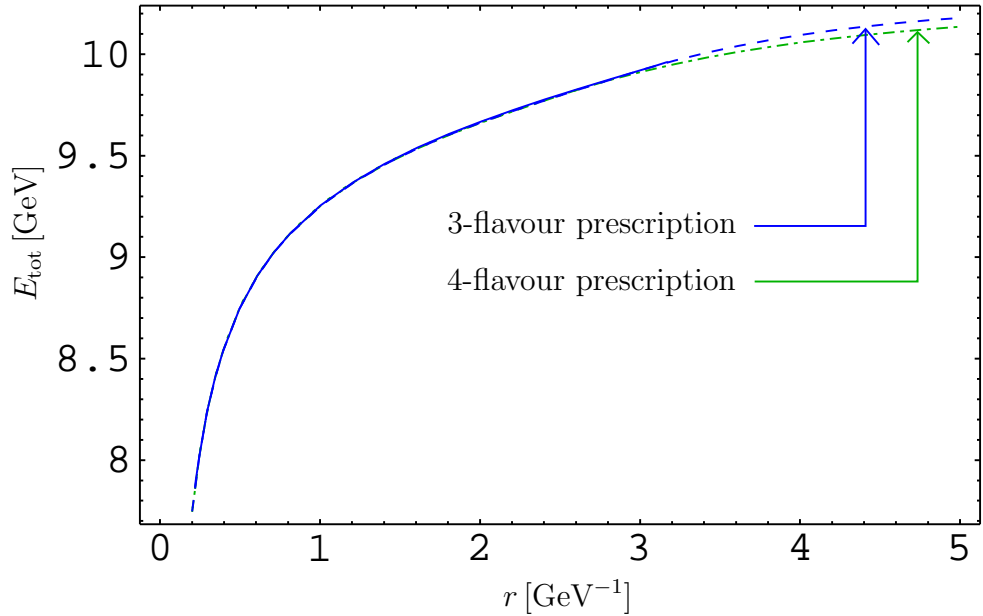}
\end{center}
\vspace*{-.5cm}
\caption{Total energy for 3-flavour ($\mu=\mu_1$: solid, 
$\mu=\mu_2$: dashed) and 4-flavour ($\mu=\mu_1$: dotted, 
$\mu=\mu_2$: dash-dotted) case. For the 3- (4-) flavour case 
prescription 1 breaks down at $r\approx 3.2\,(2.75)~{\rm GeV}^{-1}$. 
Where both scale prescriptions exist, the curves coincide. 
$\alpha_S^{(5)}(M_Z)=0.1181$.}
\label{4curves}
\end{figure}

First we examine the scale dependences of $E_{\rm tot}$ and 
$\Bigl[ E^{[3]}_{\rm tot}(r;\overline{m}_i,\alpha_S(\mu)) \Bigr]^2$
in Eqs.~(\ref{scalefix1}) and (\ref{scalefix2}).
In the first scale-fixing prescription,
the minimal sensitivity scale $\mu_1(r)$ exists only in
the range $r \simlt 3$~GeV$^{-1}$.
In the second prescription, $\mu=\mu_2(r)$,
the minimum value of $|E^{[3]}_{\rm tot}(r)|$ is zero in the range
$r \simlt 3$~GeV$^{-1}$, whereas $|E^{[3]}_{\rm tot}(r)|>0$
in the range $r \simgt 3$~GeV$^{-1}$.
These features indicate an instability of the perturbative
prediction for $E_{\rm tot}(r)$ at $r \simgt 3$~GeV$^{-1}$.
Qualitative features of the scale dependences are similar for the expansions
in the 3-flavour and the 4-flavour couplings.
(They are also similar to
the results of the analysis for $m_c\to 0$ or $m_c \to m_b$ \cite{paper1}.)
The range where the prediction is stable extends to slightly longer
distances for the 3-flavour case, in accord with our naive expectation.
The scale-dependence is demonstrated in Fig.~\ref{findmu} 
for $r=2~{\rm GeV}^{-1}$
and for the expansion in the 3-flavour coupling.
We compare the total energy determined in the four different
prescriptions (scale-fixing prescriptions 1 and 2, and expansions in
$\alpha_S^{(3)}$ and $\alpha_S^{(4)}$) in Fig.~\ref{4curves}.
All four curves agree 
well in the range where both scale-fixing prescriptions exist.
The numerical values of the total energy and the scale are listed in
Table~\ref{upsilon}.
\begin{table}
\begin{center}
\begin{tabular}{|c|c|c|}
\hline
$r~[{\rm GeV}^{-1}]$ & $E_{\rm tot}(r)$~[GeV] & $\mu_2(r)$~[GeV] \\
\hline
 0.5 & 8.75 & 1.99 \\
 1.0 & 9.25 & 1.52 \\
 1.5 & 9.49 & 1.28 \\
 2.0 & 9.66 & 1.12 \\
 2.5 & 9.79 & 0.98 \\
 3.0 & 9.92 & 0.79 \\
\hline
\end{tabular}
\caption{Comparison of the total energies and scales for the $b\bar{b}$
system (3--flavour prescription, $\mu=\mu_2$, $\alpha_S^{(5)}(M_Z)=0.1181$).
}
\label{upsilon}
\end{center}
\end{table}
We observe a strong cancellation of the leading renormalon between
the pole mass and the QCD potential: the expansion of $E_{\rm tot}(r)$
is much more convergent than the expansions of the individual terms.
This is demonstrated in Table~\ref{cancellation1}.
\begin{table}
\begin{center}
\begin{tabular}{|c|cccc|c|}
\hline
 $n$ & $m^{[n]}_{b,{\rm pole}}$ & $\delta m_{b, {\rm pole}}^{[n]}$ &
   $V_{\rm QCD}^{[n]}$ & $\delta V_{\rm QCD}^{[n]}$ & $E_{\rm tot}^{[n]}$  \\
\hline
0 & 4.190 & 0 & 0 & 0 & 8.380 \\ 
1 & 0.657 & 0 & $-0.493$ & 0 & 0.822 \\ 
2 & 0.175 & 0.028 & $-0.298$ & $-0.032$ & 0.076 \\ 
3 & 0.182 & 0.044 & $-0.391$ & $-0.084$ & $-0.024$ \\
\hline
\end{tabular}
\caption{Renormalon cancellation in the $b\bar{b}$ system
 ($r=1~{\rm GeV}^{-1}$, $\mu=\mu_1$, 3--flavour prescription, $\alpha_S^{(5)}(M_Z)=0.1181$): 
The $\alpha_S$--expansion of $E_{\rm tot}$ 
has much better convergence properties than the expansions of the
individual terms. 
$X^{[n]}$ represents the order $(\alpha_S^{(3)}(\mu ))^n$ 
term of $X$.
$m_{b, {\rm pole}}$ and $V_{\rm QCD}$ do not contain
the $m_c$--effects, these are given as $\delta m_{b, {\rm pole}}$ and
$\delta V_{\rm QCD}$.
The units are GeV.}
\label{cancellation1}
\end{center}
\end{table}

\begin{figure}
\begin{center}
\psfrag{XXX}{$r\,[{\rm GeV}^{-1}]$}
\psfrag{YYY}{$E_{\rm tot}\,[{\rm GeV}]$}
\psfrag{mc0}{$m_c \to 0$}
\psfrag{mcmb}{$m_c \to m_b$}
\psfrag{mcreal}{realistic $m_c$}
\includegraphics[width=13cm]{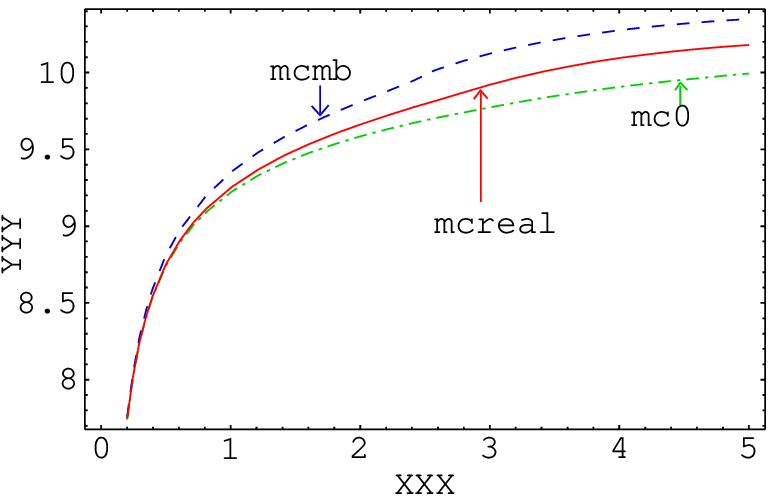}
\end{center}
\vspace*{-.5cm}
\caption{Charm mass effects: Comparison of the QCD potential
for realistic $m_c$ with the two limiting cases $m_c \to 0$
(1 heavy flavour, 4 light flavours) and $m_c \to m_b$
(2 heavy flavours, 3 light flavours) for $\alpha_S^{(5)}(M_Z)=0.1181$.}
\label{mc0andmcmb}
\end{figure}

Since the new aspect of our analysis is the inclusion of the charm-mass
effects, we compare our result with those of the analysis done in the two
hypothetical cases $m_c \to 0$ and $m_c \to m_b$.
This is shown in Fig.~\ref{mc0andmcmb}.
Our curve lies between those of the two hypothetical cases, which is consistent
with our expectation.

\begin{figure}
\begin{center}
\psfrag{XXX}{$r\,[{\rm GeV}^{-1}]$}
\psfrag{YYY}{$E\,[{\rm GeV}]$}
\psfrag{PowerLaw}{Power--law potential}
\psfrag{Log}{Log potential}
\psfrag{Cornell}{Cornell potential}
\psfrag{1161}{$\alpha_s(M_Z)=0.1161$}
\psfrag{1181}{$\alpha_s(M_Z)=0.1181$}
\psfrag{1201}{$\alpha_s(M_Z)=0.1201$}
\includegraphics[width=13cm]{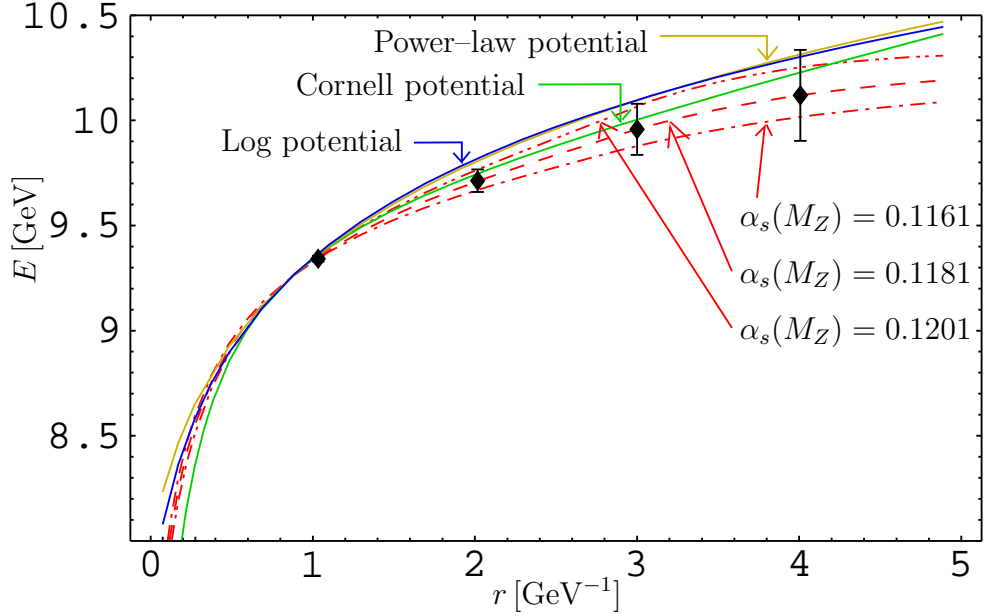}
\end{center}
\vspace*{-.5cm}
\caption{Comparison of the QCD potential to various models.
The QCD potential is given for three values of
$\alpha_S^{(5)}(M_Z)$, constants are added to make all curves coincide
with the QCD potential ($\alpha_S^{(5)}(M_Z)=0.1181$) at $r=1~{\rm GeV}^{-1}$.}
\label{comparemodels}
\end{figure}

Now we compare the total energy $E_{\rm tot}(r)$, as obtained above,
with typical phenomenological potentials in the literature:
\begin{itemize}
\item
A Coulomb-plus-linear potential (Cornell potential) \cite{cornell}:
\bea
V(r) = - \frac{\kappa}{r} + \frac{r}{a^2}
\eea
with $\kappa = 0.52$ and $a = 2.34$~GeV$^{-1}$.
\item
A power-law potential \cite{martin}:
\bea
V(r) = - 8.064~{\rm GeV} +
(6.898~{\rm GeV})(r\times 1~{\rm GeV})^{0.1} .
\eea
\item
A logarithmic potential \cite{qr}:
\bea
V(r) = -0.6635~{\rm GeV} + (0.733~{\rm GeV}) \log (r\times 1~{\rm GeV}) .
\eea
\end{itemize}
In Fig.~\ref{comparemodels} we compare these potentials with the
perturbative computation of $E_{\rm tot}(r)$ for the input values
$\alpha_S^{(5)}(M_Z)=0.1181 \pm 0.0020$.
An arbitrary constant is added to each potential, as well as
to $E_{\rm tot}(r)$, such that it coincides 
with $E_{\rm tot}(r)$ for $\alpha_S^{(5)}(M_Z)=0.1181$ at 
$r=1~{\rm GeV}^{-1}$.\footnote{
The choice of $r$, where the potentials are made to coincide, 
is to some extent arbitrary.  
(Our choice $r=1~{\rm GeV}^{-1}$ corresponds approximately 
to the root-mean-square 
radius of the $\Upsilon(1S)$ state, the heaviest state used
to determine the phenomenological potentials.)
The important point here is that we can 
indeed choose an additive constant such that $E_{\rm tot}(r)$ can be made 
to agree with the phenomenological potentials inside the 
estimated errors.
}
This is because the constant part ($r$-independent part) of each
phenomenological potential is not determined well.
As can be seen,  $E_{\rm tot}(r)$ and the phenomenological potentials
agree inside the errors estimated from the next-to-leading
renormalons $\pm \frac{1}{2} \Lambda_{\rm QCD} (\Lambda_{\rm QCD}\cdot r)^2$
(taking $\Lambda_{\rm QCD}=300$~MeV, indicated by error-bars), and
the agreement is better for a larger value of the input 
parameter $\alpha_S^{(5)}(M_Z)$.

\begin{figure}
\begin{center}
\psfrag{Delta0}{\begin{large} $\Delta_{n=0}$ \end{large}}
\psfrag{alphas}{$\alpha_S^{(5)}(M_Z)=$}
\includegraphics[width=13cm]{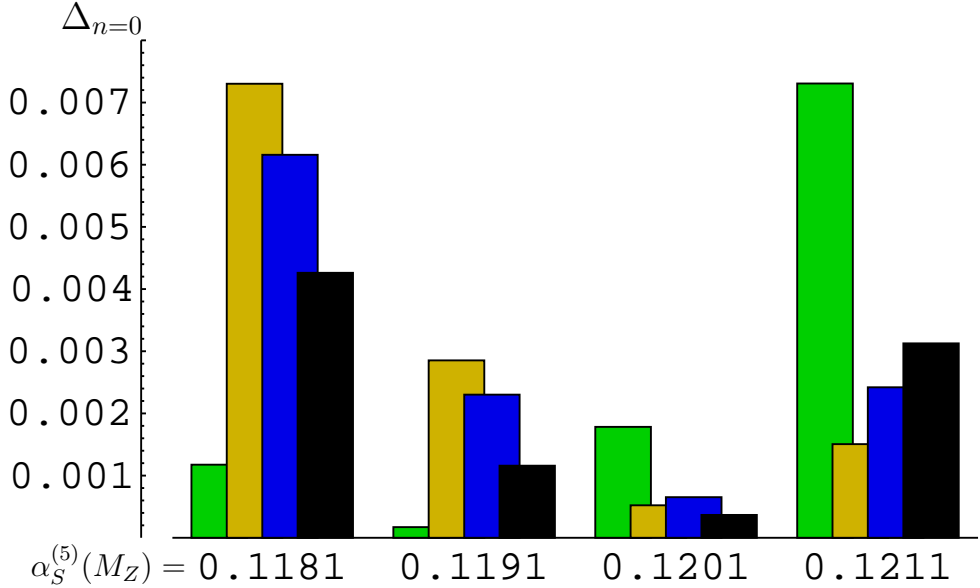}
\end{center}
\vspace*{-.5cm}
\caption{A comparison of the QCD potential for different $\alpha_S^{(5)}(M_Z)$
to various models. The height of the bars indicate the minimised
difference $\Delta_{n=0}$
between the QCD potential and the model when varying
the constant term. 
The four bars in each group correspond to
(from left to right): Cornell potential, power--law potential, 
log potential and the average of the former three.}
\label{barchart}
\end{figure}

In order to quantify the differences 
between the phenomenological potentials and the QCD potential for
different values of $\alpha_S(M_Z)$,
%of the potentials, 
we define a weighted
difference between $E_{\rm tot}(r)$ and a potential $V(r)$ as
\bea
\Delta_n [E_{\rm tot}(r) - V(r)] =
\min_{c} \,
\int^{r_1}_{r_0} dr \, r^n \, |E_{\rm tot}(r) - V(r) - c |^2 .
\eea
The minimum value of the integral is taken
as we vary an arbitrary constant added to the potential.
We examined the difference $\Delta_n$ between $E_{\rm tot}(r)$ and each of the 
phenomenological potentials, as well as $\Delta_n$ 
between $E_{\rm tot}(r)$ and the average of the three phenomenological 
potentials, for $n=-1$, 0 and $+1$.
We found similar qualitative features:
$\Delta_n$ becomes small for a larger value of
$\alpha_S^{(5)}(M_Z)$, inside the present error, around 
0.1191--0.1201.
We demonstrate this in Fig.~\ref{barchart}:
$\Delta_{n=0}$ with $r_0=1~{\rm GeV}^{-1}$ and $r_1=4~{\rm GeV}^{-1}$
is shown for each of the potentials and for the averaged
potential, varying the value of $\alpha_S(M_Z)$.

\subsection{Error Estimates}
\label{errorest}

There are several uncertainties in our theoretical prediction
for the total energy $E_{\rm tot}^{b\bar{b}}(r)$.
\begin{itemize}
\item[(i)]
\underline{\it Errors of the input parameters $\alpha_S^{(5)}(M_Z)$,
$\overline{m}_b$ and $\overline{m}_c$}:\\
\hspace{0.5cm}
In Fig.~\ref{comparemodels} the dependence of $E_{\rm tot}(r)$ on
$\alpha_S(M_Z)$ is shown.  
As $\alpha_S(M_Z)$ increases, $E_{\rm tot}(r)$ becomes steeper due to an
increase of the interquark force \cite{paper1}.
Variation of $E_{\rm tot}$ at $r=3~{\rm GeV}^{-1}$ is about $\pm 90$~MeV
with respect to its value at $r=1~{\rm GeV}^{-1}$, corresponding to
a variation $\alpha_S^{(5)}(M_Z)=0.1181 \pm 0.0020$.
The present uncertainties of the $b$-quark and $c$-quark $\overline{\rm MS}$ 
masses are about $\pm 30$~MeV \cite{bsv2} 
and $\pm 100$~MeV \cite{ej,bsv1}, respectively, 
and also their central values are correlated with
the value of $\alpha_S(M_Z)$.
The dependence of $E_{\rm tot}(r)$ on $\overline{m}_b$ for
small variations of
$\overline{m}_b$ is practically a
shift of $E_{\rm tot}(r)$ by $2 \Delta\overline{m}_b$
(independent of $r$).
We have confirmed this feature up to a larger variation of 
$\Delta\overline{m}_b = \pm 60$~MeV, corresponding to more conservative
error estimates in the literature.
The dependence of $E_{\rm tot}(r)$ on $\overline{m}_c$ is weak and negligible
($|\Delta E_{\rm tot}(r)| \simlt 10$~MeV for 
$\Delta\overline{m}_c = \pm 100$~MeV), since the charm 
mass effects are ${\cal O}(\alpha_S^2)$ and beyond.
Hence, of these
only the dependence on $\alpha_S(M_Z)$ matters in the comparison
of $E_{\rm tot}(r)$ and the phenomenological potentials.
\item[(ii)]
\underline{\it Unknown higher-order corrections}:\\
\hspace{0.5cm}
We may estimate the higher-order uncertainties from the size of the
next-to-leading renormalon \cite{al,paper1}.
We show its typical size 
$\pm \frac{1}{2} \Lambda_{\rm QCD} (\Lambda_{\rm QCD}\cdot r)^2$
by error-bars in Fig.~\ref{comparemodels}
taking $\Lambda_{\rm QCD}=300$~MeV.
We have also checked these estimates by computing the ${\cal O}(\alpha_S^4)$
correction to $E_{\rm tot}(r)$ in the large-$\beta_0$ approximation.
The corrections from the ${\cal O}(\alpha_S^4)$-term
%in $E_{\rm tot}(r)$ 
are 7, 47, 68~MeV at
$r=1$, 2, $3~{\rm GeV}^{-1}$, respectively.
\item[(iii)]
\underline{\it Approximations in $\delta m_{b,{\rm pole}}$}:\\
\hspace{0.5cm}
In Sec.~\ref{numanal} the charm mass
corrections to the pole mass,
$\delta m^{[2]}_{b,{\rm pole}}$ and $\delta m^{[3]}_{b,{\rm pole}}$,
have been computed in the linear approximation.
We estimate uncertainties induced by this approximation by using the full
correction Eq.~(\ref{dmpole1}) for $\delta m^{[2]}_{b,{\rm pole}}$
instead, while keeping $\delta m^{[3]}_{b,{\rm pole}}$ in the
linear approximation.
We find that $E_{\rm tot}(r)$ varies by $-7$, $-10$, $-17$~MeV at
$r=1$, 2, $3~{\rm GeV}^{-1}$, respectively.
\item[(iv)]
\underline{\it Non-perturbative effects}:\\
\hspace{0.5cm}
Up to now there exists no reliable way to estimate entire non-perturbative
effects on the QCD potential accurately \cite{bali}.
Moreover, in principle, the size of non-perturbative effects will depend on
the specific perturbative scheme used in the computation of the 
QCD potential or the total energy.
Here, we do not attempt to compute non-perturbative effects 
separately, but rather consider the differences between our predictions
(with respective perturbative errors) and the phenomenological potentials
as orders of magnitude estimates of non-perturbative effects in our
computational scheme.
From our present results, 
we may consider that non-perturbative effects would be
absorbed into perturbative uncertainties of our predictions.
This is consistent with the observation of \cite{bsv2}.
\end{itemize}
Besides, when we compare our theoretical predictions with
phenomenological potentials,
we should take into account the following uncertainty:
\begin{itemize}
\item[(v)]
\underline{\it Phenomenological Potentials vs.\ $E^{b\bar{b}}_{\rm tot}(r)$}:\\
\hspace{0.5cm}
Our ultimate goal would be to compare the exact QCD potential with the
perturbative prediction (after subtracting renormalons).
The phenomenological potentials are not direct physical observables
but are determined under certain model assumptions.
Since separation of the QCD potential from the rest of the
interactions (e.g. $1/r^2$ potential, relativistic corrections) is not
absolutely clear in the phenomenological approaches, there may be
substantial corrections between the phenomenological potentials and the
QCD potential or the total energy $E^{b\bar{b}}_{\rm tot}(r)$.
The relation can only be clarified by detailed comparisons of the predictions
of perturbative QCD with the experimental data
for various physical observables of the bottomonium states,
such as the energy spectrum, decay rates, level-transition rates, etc.
(Studies on the energy levels have been initiated in \cite{bsv1,bsv2}).
These detailed comparisons, however, 
are beyond the scope of the present paper.
\end{itemize}

\section{Other Systems}
\setcounter{footnote}{0}

Let us also examine the total energies of other quark-antiquark systems.
Theoretically we may expect that a limit of the potential energy exists 
when we send the masses of quark and antiquark to infinity.
Empirically both the bottomonium and charmonium
spectra can be reproduced well with the same phenomenological potential.
Thus, it would be interesting to examine whether the total energies
for different systems coincide up to an additive constant.

\begin{figure}
\begin{center}
\psfrag{XXX}{$r\,[{\rm GeV}^{-1}]$}
\psfrag{YYY}{$E_{\rm tot}\,[{\rm GeV}]$}
\psfrag{Bc3flavour}{$B_c$, 3 flavours}
\psfrag{Bc4flavour}{$B_c$, 4 flavours}
\psfrag{Upsilon}{$\Upsilon$}
\includegraphics[width=12.7cm]{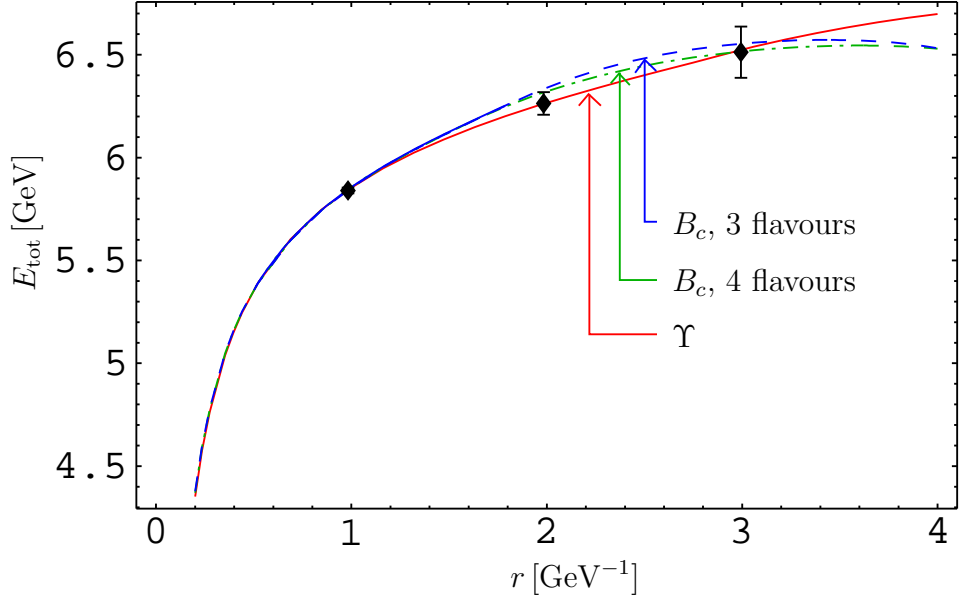}
\end{center}
\vspace*{-.5cm}
\caption{Total energy of the $B_c$--system for 3-flavour 
($\mu=\mu_1$: solid, $\mu=\mu_2$: dashed) and 4-flavour 
($\mu=\mu_1$: dotted, $\mu=\mu_2$: dash-dotted) prescription. 
%Where both scale prescriptions exist, the curves coincide.
For comparison the energy of the $\Upsilon$--system (3 flavours) 
is shown (shifted so that it coincides with the $B_c$ energy 
at $r=1~{\rm GeV}^{-1}$). For all curves $\alpha_S^{(5)}(M_Z)=0.1181$.
Error-bars indicate 
$\pm \frac{1}{2} \Lambda_{\rm QCD} (\Lambda_{\rm QCD}\cdot r)^2$
with $\Lambda_{\rm QCD}=300$~MeV.
}
\label{4bcand1upsilon}
\end{figure}

For the $b\bar{c}$ system, we consider
\bea
E_{\rm tot}^{b\bar{c}}(r) = 
m_{b,{\rm pole}} + \delta m_{b,{\rm pole}} + m_{c,{\rm pole}} 
+ V_{{\rm QCD},4}(r) + \delta V_{\rm QCD}(r) ,
\eea
where 
\bea
m_{c,{\rm pole}} = \overline{m}_c
\left\{ 1 + {4\over 3}\,  
{\alpha_S^{(3)}(\overline{m}_c)\over \pi} 
+   \left({\alpha_S^{(3)}(\overline{m}_c)\over \pi}\right)^2 d_1^{(3)} 
+  \left({\alpha_S^{(3)}(\overline{m}_c)\over \pi}\right)^3 d_2^{(3)} \right\}
.
\eea
We examine the expansions of $E_{\rm tot}^{b\bar{c}}(r)$ both in the
4-flavour and 3-flavour couplings, and in the two scale-fixing prescriptions
Eqs.~(\ref{scalefix1}) and (\ref{scalefix2}).
We find that, at distances 
$0.5~{\rm GeV}^{-1} \simlt r $ $\simlt 1.8~{\rm GeV}^{-1}$,
$\mu_1(r)$ exists in the first prescription and the minimim value of
$|E_{\rm tot}^{[3]}|$ is zero in the second prescription,
indicating that stable theoretical predictions for the total energy 
are obtained in this region.
A comparison in Fig.~\ref{4bcand1upsilon} shows
that the predictions for the total energies of the $b\bar{c}$ system 
and $b\bar{b}$ system 
agree within the uncertainties of the predictions.
We also note that, even at $r \simgt 1.8~{\rm GeV}^{-1}$,
where only the second scale-fixing prescription exists,
the curves for the $b\bar{c}$ system in the 4-flavour and 3-flavour couplings
agree with each other and
they are also consistent with that of the $b\bar{b}$ system.
See also Tables~\ref{bc} and \ref{cancellation2} for numerical values.
\begin{table}
\begin{center}
\begin{tabular}{|c|c|c|}
\hline
$r~[{\rm GeV}^{-1}]$ & $E_{\rm tot}(r)$~[GeV] & $\mu_2(r)$~[GeV] \\
\hline
 0.5 & 5.35 & 4.08 \\
 1.0 & 5.85 & 1.31 \\
 1.5 & 6.12 & 0.97 \\ 
 2.0 & 6.34 & 0.76 \\
 2.5 & 6.48 & 0.83 \\ 
 3.0 & 6.55 & 0.93 \\
\hline
\end{tabular}
\caption{Comparison of the total energies and scales for the $b\bar{c}$
system (3--flavour prescription, $\mu_2$).
The predictions are stable in the range
$0.5~{\rm GeV}^{-1} \simlt r \simlt 1.8~{\rm GeV}^{-1}$.
}
\label{bc}
\end{center}
\end{table}
\begin{table}
\begin{center}
\begin{tabular}{|c|ccccc|c|}
\hline
 $n$ & $m^{[n]}_{b,{\rm pole}}$ & $\delta m_{b, {\rm pole}}^{[n]}$ 
& $m_{c, {\rm pole}}^{[n]}$ &
   $V_{\rm QCD}^{[n]}$ & $\delta V_{\rm QCD}^{[n]}$ & $E_{\rm tot}^{[n]}$  \\
\hline
0 & 4.190 & 0     & 1.243 &   0      &   0      &   5.433 \\
1 & 0.730 & 0     & 0.217 & $-0.547$ &   0      &   0.399 \\
2 & 0.153 & 0.035 & 0.211 & $-0.321$ & $-0.039$ &   0.038 \\
3 & 0.203 & 0.054 & 0.304 & $-0.466$ & $-0.109$ & $-0.014$ \\
\hline
\end{tabular}
\caption{Renormalon cancellation in the $b\bar{c}$ system
 ($r=1~{\rm GeV}^{-1}$, $\mu=\mu_1$, 3--flavour prescription).
Notations are same as in Table~\ref{cancellation1}.
}
\label{cancellation2}
\end{center}
\end{table}

Furthermore we examined the total energy of the $c\bar{c}$ system,
\bea
E_{\rm tot}^{c\bar{c}}(r) = 2 m_{c,{\rm pole}} + V_{\rm QCD,3}(r) ,
\eea
in a similar
way.
We, however, obtained stable predictions only in the very narrow
vicinity of $r \simeq 1~{\rm GeV}^{-1}$ and consequently we could not
compare the shape of $E_{\rm tot}(r)$ with that of the $b\bar{b}$ or
$b\bar{c}$ system.
We believe that this is caused by the typical scales for the $c\bar{c}$
system being too low to give reliable perturbative expansions.
(We may compare this feature with that of the perturbative calculation
of the energy levels of charmonium \cite{bsv1}:
only the $1S$ levels can be computed reliably.)

\begin{figure}
\begin{center}
\psfrag{XXX}{$r\,[{\rm GeV}^{-1}]$}
\psfrag{YYY}{$E_{\rm tot}\,[{\rm GeV}]$}

\psfrag{3flavour}{$\tilde \Upsilon$, 3 flavours}
\psfrag{4flavour}{$\tilde \Upsilon$, 4 flavours}
\psfrag{Upsilon}{$\Upsilon$}
\includegraphics[width=13cm]{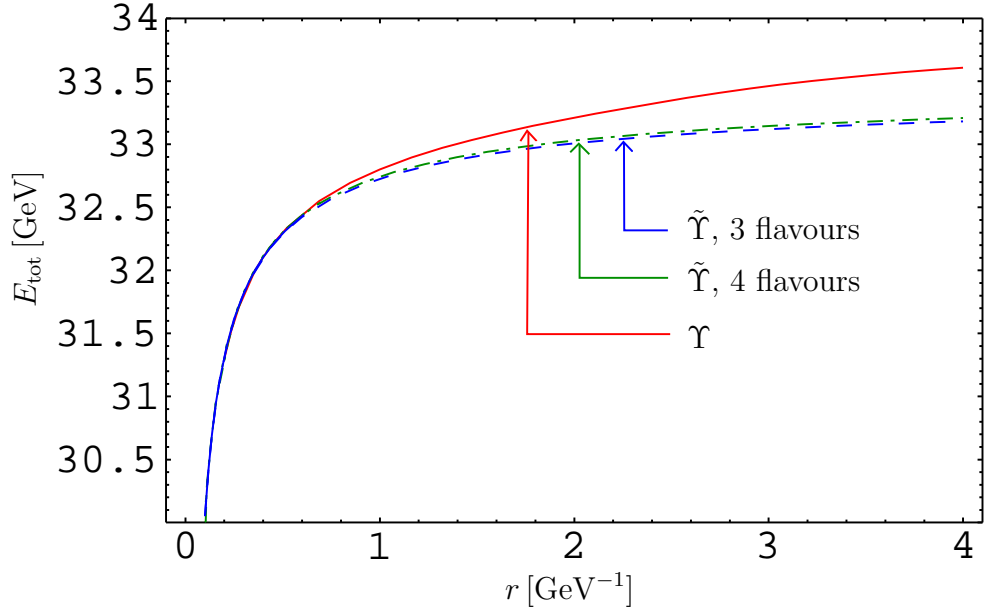}
\end{center}
\vspace*{-.5cm}
\caption{Total energy of the bound state $\tilde \Upsilon$ of a hypothetical
heavy $b$--quark with a mass of 15 GeV in 3-flavour
($\mu=\mu_1$ [breaks down at $r\approx 0.6\,{\rm GeV}^{-1}$]: solid,
 $\mu=\mu_2$: dashed) and four flavour
($\mu=\mu_1$ [breaks down at $r\approx 0.8\,{\rm GeV}^{-1}$]: dotted, 
 $\mu=\mu_2$: dash-dotted) prescription.
Where both scale prescriptions exist, the curves coincide.
For comparison the energy for the $\Upsilon$--system (3 flavours)
is shown (shifted so that it coincides with the $\tilde \Upsilon$ energy
at $r=0.5\,{\rm GeV}^{-1}$. 
For all curves $\alpha_S^{(5)}(M_Z)=0.1181$.
}
\label{heavyb}
\end{figure}

Conversely we 
examined the total energy of the $b\bar{b}$ system as we increase
the bottom quark mass artificially.
When we do this, we find that the region of $r$, 
where we can make a stable prediction of $E_{\rm tot}(r)$, shifts to shorter 
distances.  
In other words, we can still make a stable prediction of 
the shape of the potential from the distance $r \simeq 1/\overline{m}_b$ 
up to the region which is relevant to the 
formation of the (hypothetically heavy) bottomonium states.
This is reasonable, since a heavier quarkonium 
has a smaller radius, and the prediction should be stable for this smaller
radius.
On the other hand, we do not always have stable predictions 
for $E_{\rm tot}(r)$ in the region of 
our interest, $0.5~{\rm GeV}^{-1} \simlt r \simlt 5~{\rm GeV}^{-1}$.  
We may understand this property as a manifestation of a multi-scale problem.  
The pole mass contains powers of $\log(\mu/\overline{m}_b)$,
while the QCD potential contains powers of $\log(\mu r)$.  
When $\overline{m}_b$ and $r^{-1}$ are very different, 
it generally becomes more difficult to cancel 
two different types of logarithms, so it becomes difficult to find a scale 
$\mu$ which stabilizes the theoretical prediction.
For a (hypothetical) value of $\overline{m}_b$ below about $23$~GeV,
we still have an overlap between the region of
stable predictions for the total energy and the above region of our interest.
Where both scale-fixing prescriptions exist, the predictions 
coincide.
(The distance $r$ where the
minimum of $|E_{\rm tot}^{[3]}|$ deviates from zero in the second
prescription is close to the distance beyond which the
first scale $\mu_1(r)$ does not exist.)
In the region where both prescriptions exist, the predictions agree with 
$E^{b\bar{b}}_{\rm tot}(r)$ computed with the realistic
$b$-quark mass, up to an additive 
constant.\footnote{
Of course, the agreement is better at shorter distances, where the
perturbative predictions tend to be more reliable.
At $r\simlt 1.5~{\rm GeV}^{-1}$ the expansion in the 4-flavour coupling
tends to be more stable than the expansion in the 3-flavour coupling.
}
At larger $r$ the larger mass produces a smaller total energy. 
This tendency is expected from the higher-order analysis in the
large-$\beta_0$ approximation \cite{paper1}.
Also this agrees qualitatively 
with our results for the $b\bar{c}$ against $b\bar{b}$ system:
from the point where the curves split, $E_{\rm tot}^{b\bar{c}}$ 
is larger than $E_{\rm tot}^{b\bar{b}}$, although the curves later cross
(Fig.~\ref{4bcand1upsilon}).
These features are demonstrated in Fig.~\ref{heavyb}, where the total energy
for $\overline{m}_b=15$~GeV is shown and compared with the one with the
realistic $b$-quark mass.

\section{Conclusions}

We have analysed the total energy of the $b\bar{b}$ system
incorporating the non-zero charm-quark mass effects.
We observed an improvement of
convergence of the perturbative expansion, once we perform the cancellation of
the leading renormalons;
we obtained stable theoretical predictions for the total energy
$E_{\rm tot}^{b\bar{b}}(r)$ at $r \simlt 3~{\rm GeV}^{-1}$.
These features are qualitatively the same as those observed in \cite{paper1}.
We compared the total energy with typical phenomenological potentials.
They agree in the range $0.5~{\rm GeV}^{-1} \simlt r \simlt 3~{\rm GeV}^{-1}$ 
within the estimated theoretical uncertainties.
The agreement becomes very good
when the value of the input parameter
$\alpha_S^{(5)}(M_Z)$ is large (inside the error bands),
0.1191--0.1201.
From these results, we may conclude that, in analysing the nature
of the bottomonium states, using the perturbative prediction for
$E_{\rm tot}^{b\bar{b}}(r)$ is, at least, as good as using 
phenomenological potentials.
We may compare our result with the comprehensive
analysis of the bottomonium spectrum which includes full
corrections up to ${\cal O}(1/c^2)$ \cite{bsv2}.
There, a smaller value around 0.1161 for $\alpha_S^{(5)}(M_Z)$
was favored.
Therefore, we see that 
the interactions other than $E_{\rm tot}(r)$ play non-negligible 
roles for the predictions of the bottomonium energy levels, with respect 
to the present theoretical accuracy of perturbative QCD.

We also examined the perturbative predictions for
$E_{\rm tot}(r)$ of other systems.
For the $b\bar{c}$ system, stable predictions are obtained (at least) in the
range $r \simlt 1.8~{\rm GeV}^{-1}$.
In this range, the total energy agrees with that of the $b\bar{b}$ system
inside theoretical uncertainties.
For the $c\bar{c}$ system, we could barely obtain stable theoretical
predictions for the shape of $E_{\rm tot}(r)$, since the relevant scales
are very low.
We also found that, for a heavier (hypothetical)
quarkonium system, the range where
$E_{\rm tot}(r)$ can be predicted reliably, shifts to shorter distances.
At any event,
whenever we may obtain stable theoretical predictions for $E_{\rm tot}(r)$
at $r \simgt 0.5~{\rm GeV}^{-1}$, the predictions  
agree with the phenomenological potentials within present theoretical
uncertainties.

\section*{Appendix}

\subsection*{A: Parameters}

The constants used in Eqs.~(\ref{massrel}) and (\ref{QCDpot}) 
are given by \cite{ps}
\bea
&&
a_1^{(n_l)} = \frac{31}{3} - \frac{10}{9} n_l ,
\\ &&
a_2^{(n_l)} = {\frac{4343}{18}} + 36\,{{\pi }^2} +   66\,{\zeta_3} - 
  {\frac{9\,{{\pi }^4}}{4}} - {n_l}\,
   \left( {\frac{1229}{27}} + {\frac{52\,{\zeta_3}}{3}} \right)  
+ {\frac{100}{81}} \,{n_l^2} ,
\nonumber \\
\eea
and
\bea
d^{(n_l)}_1 &=&
{\frac{307}{32}} + {\frac{{{\pi }^2}}{3}} 
      + {\frac{{{\pi }^2}\,\log 2}{9}} - {\frac{\zeta_3}{6}}
+ 
  n_l\,\left( -{\frac{71}{144}} - {\frac{{{\pi }^2}}{18}} \right)
\nonumber \\ &\simeq&
13.4434-1.04137 \, n_l ,
\\ ~~~ \nonumber \\
d^{(n_l)}_2 &=&
{\frac{8462917}{93312}} + {\frac{652841\,{{\pi }^2}}{38880}} - 
  {\frac{695\,{{\pi }^4}}{7776}} - {\frac{575\,{{\pi }^2}\,\log 2}{162}} 
\nonumber \\ &&
- 
  {\frac{22\,{{\pi }^2}\,{{\log^2 2}}}{81}} - 
  {\frac{55\,{{\log^4 2}}}{162}} 
- 
  {\frac{220\,{\rm Li}_4(\frac{1}{2})}{27}} 
+ {\frac{58\,\zeta_3}{27}} - 
  {\frac{1439\,{{\pi }^2}\,\zeta_3}{432}} + 
  {\frac{1975\,\zeta_5}{216}}
\nonumber \\ &&
+ 
  n_l\,\left( -{\frac{231847}{23328}} - 
     {\frac{991\,{{\pi }^2}}{648}} + {\frac{61\,{{\pi }^4}}{1944}} - 
     {\frac{11\,{{\pi }^2}\,\log 2}{81}} + 
     {\frac{2\,{{\pi }^2}\,{{\log^2 2}}}{81}} + 
     {\frac{{{\log^4 2}}}{81}} 
\right.
\nonumber \\ &&
\left.
+ 
     {\frac{8\,{\rm Li}_4(\frac{1}{2})}{27}} - 
     {\frac{241\,\zeta_3}{72}} \right)  + 
  {n_l^2}\,\left( {\frac{2353}{23328}} + 
     {\frac{13\,{{\pi }^2}}{324}} + {\frac{7\,\zeta_3}{54}}
     \right)  
\nonumber \\ &\simeq &
190.391 - 26.6551\, n_l +  0.652691\, n_l^2 .
\eea

%\newpage
\subsection*{B: \boldmath{$\delta V_{\rm QCD}^{[3]}(r)$}}

After correcting misprints in \cite{hoangmceff,melles2}, 
the charm-mass correction
to the QCD potential at ${\cal O}(\alpha_S^3)$ reads\footnote{
The last line stems from our use of $\overline{m}_c$ instead of
the pole mass, and it is not due to the misprint.
}
\bea
\delta V_{\rm QCD}^{[3]}(r) &=&
- \, \frac{4}{3} \,  \frac{\alpha_S^{(4)}(\mu)}{r}
\biggl( \frac{\alpha_S^{(4)}(\mu)}{3\pi} \biggr)^2 
\Biggl[ 
\nonumber \\ &&
\Biggl\{
- \frac{3}{2} \int_1^\infty dx \, f(x) \, e^{-2 \overline{m}_c r x}
\Bigl( \beta_0^{(4)} \Bigl( \log \frac{4\overline{m}_c^2x^2}{\mu^2} 
- {\rm Ei}(2\overline{m}_crx) - e^{4 \overline{m}_c r x}{\rm Ei}(-2\overline{m}_crx) \Bigr) - a_1^{(4)} \Bigr)
\nonumber\\ &&
+3 \, \Bigl( \log (\overline{m}_c r) + \gamma_E + \frac{5}{6} \Bigr) 
\Bigl( \beta_0^{(4)} \ell + \frac{a_1^{(4)}}{2} \Bigr)
+ \beta_0^{(4)} \, \frac{\pi^2}{4} \Biggr\}
\nonumber \\ 
&-& \Biggl\{ \int_1^\infty dx \, f(x) \, e^{-2 \overline{m}_c r x}
\Bigl( \frac{1}{x^2} + x \, f(x) \, \log 
\frac{x-\sqrt{x^2-1}}{x+\sqrt{x^2-1}} \Bigr)
\nonumber \\ &&
+ \int_1^\infty dx \, f(x) \, e^{-2 \overline{m}_c r x}
\Bigl( \log 4x^2 - {\rm Ei}(2\overline{m}_crx) - e^{4 \overline{m}_c r x}{\rm Ei}(-2\overline{m}_crx) \Bigr)
\nonumber \\ &&
- \Bigl( \log (\overline{m}_c r) + \gamma_E + \frac{5}{6} \Bigr)^2
- \frac{\pi^2}{12} \Biggr\}
\nonumber \\
&+& \Biggl\{
\frac{57}{4} \Bigl( f_1 \, \Gamma ( 0, 2f_2 \overline{m}_c r) 
+ b_1 \, \Gamma(0,2b_2 \overline{m}_c r) + \log (\overline{m}_c r) + \gamma_E
+ \frac{161}{228} + \frac{13}{19}\zeta_3 \Bigr) \Biggr\}
\nonumber \\
&+& 
\Biggl\{ \int_1^\infty dx \, f(x) \, e^{-2 \overline{m}_c r x}
\Bigl( - {8 \overline{m}_c r x} \Bigr) + 4 \Biggr\}
\Biggr] .
\eea

\section*{Acknowledgements}
Y.S. is grateful to Y. Kiyo for fruitful discussion.
S.R. was supported by the Japan Society for the Promotion of Science
(JSPS).

\end{document}